\documentclass[12pt]{iopart}
\usepackage{graphicx}
\DeclareGraphicsExtensions{.eps,.ps}
\begin{document}

\def\bbm[#1]{\mbox{\boldmath$#1$}}

\title[Initial correlations effects on decoherence at zero temperature]
{Initial correlations effects on decoherence at zero temperature}

\author{B. Bellomo\dag\, G. Compagno\dag\
and F. Petruccione\ddag }

\address{\dag\ Dipartimento di Scienze Fisiche ed
Astronomiche dell'Universit\`{a} di Palermo, Via Archirafi, 36,
I-90123 Palermo, Italy.}

\address{\ddag\ School of Pure and Applied Physics Howard College,
University of KwaZulu-Natal, Durban, 4041 South Africa.}

\ead{bruno.bellomo@fisica.unipa.it, compagno@fisica.unipa.it}

\begin{center}
(Dated: November 15, 2005)
\end{center}

\begin{abstract}
We consider a free charged particle interacting with an
electromagnetic bath at zero temperature. The dipole approximation
is used to treat the bath wavelengths larger than the width of the
particle wave packet. The effect of these wavelengths is described
then by a linear Hamiltonian whose form is analogous to
phenomenological Hamiltonians previously adopted to describe the
free particle-bath interaction. We study how the time dependence
of decoherence evolution is related with initial particle-bath
correlations. We show that decoherence is related to the time
dependent dressing of the particle. Moreover because decoherence
induced by the $T=0$ bath is very rapid, we make some
considerations on the conditions under which interference may be
experimentally observed.

\end{abstract}





\section{Introduction}

Decoherence, that is the destruction of coherent phase relations
present in the elements of the density matrix describing the
system, is induced by the interaction between the quantum system
and the environment. It appears to be relevant in fields ranging
from quantum measurement \cite{Libro decoerenza (2002)} to
classical-quantum interface \cite{Zurek (2003)}, quantum
information theory and computation \cite{Unruh (1995)} and
cosmology \cite{Barvinsky (1999)}. Recently experimental evidence
of environment induced decoherence has also been reported
\cite{Brune (1996),Myatt (2000),Raimond (2001), Brezger
(2002),Hornberger (2003),Auffeves (2003),Hackermüller (2004)}.
Several analysis of decoherence processes have been reported for
the case of a particle, either free or in a potential, linearly
coupled to the environment modelled as a bath of harmonic
oscillators at temperature $T$ \cite{Hakim-Ambegaokar
(1985),Barone-Caldeira (1991), Ford (1993), Durr-Spohn (2000),
Breuer-Petruccione (2001), Mazzitelli (2003)}. Usually, initial
conditions are adopted in which the system and the environment are
decoupled, the interaction being effective after some initial time
$t=0$ \cite{Feynman-Vernon (1963),Caldeira-Leggett
(1985),Privman-Mozyrsky (1998),Tolkunov-Privman (2004)}. It has
been shown that, starting with factorizable initial conditions,
which corresponds to the absence of initial correlations between
the system and the environment, it is possible to separate the
decoherence into  two characteristic parts. The first, related to
the thermal properties of the bath, has a typical development time
which in some models goes like $T^{-1}$; the second, related to
the zero point fluctuations of the oscillators of the bath, has a
characteristic development time independent of temperature
\cite{Petruccione-Breuer libro (2002)}.

The time development of decoherence is also affected by the
initial presence of correlations between the system and the
environment \cite{Hakim-Ambegaokar (1985),Barone-Caldeira
(1991),Grabert-Schramm-Ingold (1988),Smith-Caldeira
(1990),Romero-Paz (1997),Ford-Lewis-O'Connell (2001),Lutz (2003)}.
However to choose the amount of correlations present in the
initial conditions and to determine their influence on the time
development may be a delicate problem mainly when interaction with
the bath is always present as in the case of a charged particle
interacting with the radiation field. However when the initial
time is taken immediately after a fast dynamical evolution, like
immediately after a collision, it appears to be reasonable to
consider conditions where the particle and the bath are not in
complete equilibrium \cite{Compagno-Passante-Persico (1995)}, the
extreme cases being either of completely factorized or completely
correlated initial conditions. Non factorized initial conditions
have in fact been adopted for particles interacting with a thermal
bath corresponding to the condition where a position measurement
has been made on the particle once it is in equilibrium with the
bath at temperature $T$ \cite{Barone-Caldeira (1991),Romero-Paz
(1997),Lutz (2003)}. Other types of initial conditions have been
also considered with the particle subject to a potential abruptly
modified at $t=0$ or a combination of this and of the previous
ones \cite{Smith-Caldeira (1990)}. The presence of entanglement
between a  particle and a bath at zero temperature has been in
particular explicitly taken into account \cite{Ford-O'Connell
(2003)}.

In the context of quantum computing it has been shown, that in an
ensemble of two level systems interacting with a reservoir of
harmonic oscillators, decoherence among the two level systems
develops because of the buildup of correlations between each two
level system and the environment \cite{Palma-Suominen-Ekert
(1996)}. A similar mechanism has been suggested to occur also in
the case of a free charged particle interacting with the
electromagnetic field vacuum. Here the decoherence among different
momenta of the particle wave packet develops because of the
buildup of correlations between each momentum and the associated
transverse electromagnetic field structure that is responsible of
the mass renormalization \cite{bellomo (2004)}.

To treat the case of a non relativistic charged particle
interacting, within the electric dipole approximation, with the
electromagnetic field at temperature $T$, both the Hamiltonian
approach \cite{Durr-Spohn (2000)} and  functional techniques
\cite{Barone-Caldeira (1991),Breuer-Petruccione (2001)} have been
used. In particular development of decoherence has been studied
for charged particles by examining the time dependence of the
interference among two coherent wave packets.

Recently diffraction experiments have been performed where
interference among different wave packets of the same particle has
been observed also for rather long times  \cite{Brune (1996),Myatt
(2000), Brezger (2002), Hackermüller (2004)}. There are however
indications that starting from uncoupled initial conditions
decoherence develops also when the environment is at zero
temperature and usually it may occur faster than the typical times
of motion \cite{Ford (1993),Petruccione-Breuer libro (2002)}. It
appears thus of interest to examine how the development of vacuum
induced decoherence depends from the amount of correlation
initially present between the particle and the environment.

Our model consists of a non relativistic free particle linearly
interacting with a bath of harmonic oscillators at zero
temperature. Our results will in particular be specialized to the
case of a charged particle embedded in the electromagnetic field
modes represented by a set of harmonic oscillators. This
specialization is obtained from the general case by choosing the
appropriate form of the coupling constants.

To study the effects of initial condition on the time development
of decoherence among the momentum components of the same wave
packet, the behavior of the reduced density matrix elements shall
be analyzed.

\section{Model} \label{par:Modello}

Phenomenological Hamiltonians, where the interaction is described
by a linear coupling, have previously been adopted  in the study
of decoherence of free particles interacting with the environment
treated as a bath of harmonic oscillators \cite{Hakim-Ambegaokar
(1985), Ford (1993), Breuer-Petruccione (2001)}.

Here we consider a non relativistic free charged particle,
initially moving at a velocity $\bar{\bbm[v]}$, interacting with
the transverse electromagnetic field.

Taking the electromagnetic field as a set of normal modes, each
characterized by a wave vector $\bbm[k]$ and a polarization index
$j$, the potential vector in the Coulomb gauge and with periodic
boundary conditions taken on a volume $V$ is
\begin{equation}\label{potenziale vettore}
 \hat{\bbm[A]} (\hat{\bbm[r]})= \sum_{k,j}
\bbm[\varepsilon]_{k,j}\sqrt{\frac{2\pi\hbar c^2}{V \omega_k
}}\left(\hat{\mathrm{a}}^{\dag}_{k,j}\mathrm{e}^{-i
\bbm[k]\cdot\hat{\bbm[r]}}+\hat{\mathrm{a}}_{k,j}\mathrm{e}^{i
\bbm[k]\cdot\hat{\bbm[r]}} \right) \, ,
\end{equation}
where $\bbm[\varepsilon]_{k,j}$ are the polarization vectors
$(j=1,2)$, $\hat{\bbm[r]}$ is the particle position operator and
$\hat{\mathrm{a}}_{k,j}$ and $\hat{\mathrm{a}}^{\dag}_{k,j}$ are
the annihilation and creation operators of the modes that satisfy
the commutation rules
$[\hat{\mathrm{a}}_{k,j},\hat{\mathrm{a}}^{\dag}_{k',j'}]=\delta_{kk'}\delta_{jj'}$.
\\
The non relativistic minimal coupling Hamiltonian has the form
\begin{eqnarray}\label{hamiltoniana
completa}  \hat{H} =  \frac{1}{2m_0 }\left[\hat{\bbm[p]} - \frac{e
\hat{\bbm[A]}
  (\hat{\bbm[r]})}{c} \right]^2 +\sum_{k,j}\hbar \omega_k
  \hat{\mathrm{a}}^{\dag}_{k,j}\hat{\mathrm{a}}_{k,j}
   \, ,
\end{eqnarray}
where $\hat{\bbm[p]}$ is the particle momentum operator, $m_0$ is
the bare mass and $e$ is the charge of the particle.

In the case of a free particle the dipole approximation, typically
adopted for bound charges, may be also adopted if the linear
dimensions of the wave packet are small compared to the relevant
wavelengths of the field.  This can be applied to our Hamiltonian
(\ref{hamiltoniana completa}) introducing a cut off frequency
$\Omega$ such that the corresponding wavelength still allows the
application of dipole approximation. Moreover in order to adopt
this approximation without having problems related to the distance
covered in time by the particle, taking advantage of Galilean
invariance of the non relativistic Hamiltonian, we put ourselves
in the reference system comoving with the particle so that its
average initial velocity is zero \cite{Mazzitelli (2003)}. All the
quantities we have introduced are obviously relative to this
reference system. Within the dipole approximation, the operator
$\hat{\bbm[r]}$ can be replaced by a parameter $\bbm[r]_0$
indicating the wave packet position at time $t$. In absence of
interaction, in the comoving reference system, $\bbm[r]_0$ is
obviously given by the initial position of the particle.
Substituting this parameter in Eq.\,(\ref{hamiltoniana completa}),
the Hamiltonian becomes
\begin{eqnarray}\label{hamiltoniana2}
  \hat{H}=& \frac{\hat{\bbm[p]}^2}{2m_0} +\sum_{k,j}\hbar \omega_k
  \hat{\mathrm{a}}^{\dag}_{k,j}\hat{\mathrm{a}}_{k,j}  -
  \frac{e}{m_0c} \hat{\bbm[p]} \cdot \hat{\bbm[A]}
  (\bbm[r]_0)  \, ,
\end{eqnarray}
where we have neglected the quadratic interaction term. In fact
its physical origin is linked to the particle average kinetic
energy due to the vacuum fluctuations \cite{Weisskopf (1939)}, and
it is usually very small compared to the linear term. On the other
hand it can also be exactly eliminated by a canonical
transformation of the Bogoliubov Tiablikov form
\cite{Tyablikov-Bogoliubov (1967)}. This transformation has in
fact been used to eliminate an analogous quadratic interaction
term in the phenomenological Hamiltonian used to describe a free
particle coupled to a dissipative environment
\cite{Hakim-Ambegaokar (1985)}.

The advantage of using the dipole approximation and of having a
linear interaction in our system manifests itself in the fact that
the Hamiltonian can be treated exactly. Moreover the Hamiltonian
of Eq.\,(\ref{hamiltoniana2}) is formally equivalent to a model
Hamiltonian previously used, in the context of quantum computing,
to study the decoherence of an ensemble of two level systems
coupled to a reservoir of harmonic oscillators
\cite{Palma-Suominen-Ekert (1996)}. This will permit to develop a
physical analogy between these systems.

It is clear that the wave packet describing the free particle is
subject to spreading. This limits the range of time where dipole
approximation may be used \cite{Ford (1993)}. This spreading of
the wave packet has been also taken in consideration with regard
to eventual problems in the definition of decoherence
\cite{Connell (2005)}.

Using Eqs.\,(\ref{hamiltoniana2}) and (\ref{potenziale vettore})
the Hamiltonian can be written as
\begin{eqnarray}\label{hamiltoniana01}
  \fl   \hat{H}=\sum_p \frac{p^2}{2m_0}\;\hat{\bbm[\sigma]}_p+\sum_{k,j}\hbar \omega_k
  \hat{\mathrm{a}}^{\dag}_{k,j}\hat{\mathrm{a}}_{k,j}  +
  \sum_{p,k,j}\hat{\bbm[\sigma]}_p \, g_{k,j}^p\left(
  \hat{\mathrm{a}}^{\dag}_{k,j}\mathrm{e}^{-i \bbm[k]
  \cdot \bbm[r]_0}+\hat{\mathrm{a}}_{k,j}\mathrm{e}^{i \bbm[k]\cdot \bbm[r]_0}
  \right)\,,
\end{eqnarray}
where we have introduced for the momentum operator the notation
$\sum_p \bbm[p] \, \hat{\bbm[\sigma]}_p$, with
$\hat{\bbm[\sigma]}_p=|p\rangle \langle p |$ the projection
operator on the momentum eigenvalue $\bbm[p]$, while $g_{k,j}^p$
are the coupling coefficients given by
\begin{equation}\label{coefficienti di accoppiamento}
  g_{k,j}^p=-\bbm[p]\cdot
\bbm[\varepsilon]_{k,j}\frac{e}{m_{0}}\sqrt{\frac{2\pi\hbar}{V
\omega_k }}\, .
\end{equation}
Eq.\,(\ref{hamiltoniana01}) expresses the charge-field Hamiltonian
in its unrenormalized form. To obtain the renormalized form we
introduce the physical mass $m$ as
\begin{equation}\label{massa vestita}
  \frac{1}{m_0}= \frac{1}{m}\left(1+\frac{\delta
m}{m}\right)\,,
\end{equation}
where $\delta m$ represents the mass variation due to the coupling
with the bath. The renormalized form is then
\begin{eqnarray}\label{hamiltoniana}
   \fl \hat{H}=\sum_p \frac{p^2}{2m}\;\hat{\bbm[\sigma]}_p+\sum_{k,j}\hbar \omega_k
  \hat{\mathrm{a}}^{\dag}_{k,j}\hat{\mathrm{a}}_{k,j}   +
  \sum_{p,k,j}\hat{\bbm[\sigma]}_p \, g_{k,j}^p\left(
  \hat{\mathrm{a}}^{\dag}_{k,j}\mathrm{e}^{-i  \bbm[k]\cdot \bbm[r]_0}
  +\hat{\mathrm{a}}_{k,j}\mathrm{e}^{i  \bbm[k]\cdot \bbm[r]_0}  \right)+\sum_p
\frac{p^2}{2m}\frac{\delta m}{m}\;\hat{\bbm[\sigma]}_p
\nonumber \\
\end{eqnarray}
\\
The Hamiltonian under the form (\ref{hamiltoniana01}) or
(\ref{hamiltoniana}) can describe a variety of physical systems
with an appropriate choice of the coefficients $g_{k,j}^p$. In
fact a form analogous to it has been previously used, both in its
unrenormalized \cite{Durr-Spohn (2000),Breuer-Petruccione (2001)}
and renormalized form \cite{Hakim-Ambegaokar
(1985),Barone-Caldeira (1991)}, with the appropriate $g_{k,j}^p$,
to treat a particle interacting linearly with a bath.

In the following we shall keep unexplicit the expression of the
coupling coefficients $g_{k,j}^p$, so that some results will be
true not only for the charge-field interaction but in general for
any particle-bath interaction with linear coupling. The explicit
charge-field form (\ref{coefficienti di accoppiamento}) will be
only used at the end to evidence the dependence of the results
from the parameters of the system.

\section{Fully correlated states}

To treat the effects of the presence of particle-field
correlations in the initial state, we consider at first an initial
condition with the particle in complete equilibrium with the zero
point field fluctuations and localized at the position
$\bbm[r]_0$. This implies that each momentum component of the wave
packet is in equilibrium with the fluctuations of the
electromagnetic field. The momentum of the charge commutes with
the Hamiltonian (\ref{hamiltoniana}), thus the equilibrium state
of a given momentum must be an eigenstate of $\hat{H}$. In order
to find these eigenstates we diagonalize exactly $\hat{H}$ by a
canonical transformation using the unitary operators
\begin{equation}\label{operatore trasformazione}
  \hat{D}_{k,j}^p=\exp \left[ \sum_p\frac{\hat{\bbm[\sigma]}_p \, g_{k,j}^p}{\hbar\omega_k}
  \left( \hat{\mathrm{a}}^{\dag}_{k,j}\mathrm{e}^{-i  \bbm[k]\cdot \bbm[r]_0} -
  \hat{\mathrm{a}}_{k,j}\mathrm{e}^{i  \bbm[k]\cdot \bbm[r]_0} \right)\right]\, ,
\end{equation}
that act in the particle-field mode Hilbert space
$\mathcal{H}_S\otimes\mathcal{H}_F^{k,j}$.
\\
The action of the canonical transformation, induced by
$\hat{D}_{k,j}^p$ on the operators $\hat{\mathrm{a}}_{k,j}$ and
$\hat{\mathrm{a}}^{\dag}_{k,j}$, is
\begin{eqnarray}\label{trasformazione}
&\mathrm{\widetilde{a}}_{k,j}=\hat{D}_{k,j}^{p\,-1}\hat{\mathrm{a}}_{k,j}\hat{D}_{k,j}^{p}
=\hat{\mathrm{a}}_{k,j}+\sum_p\frac{\hat{\bbm[\sigma]}_p \,
g_{k,j}^p}{\hbar\omega_k} \mathrm{e}^{i \bbm[k]\cdot \bbm[r]_0}\,
, \nonumber \\ &
\mathrm{\widetilde{a}}^{\dag}_{k,j}=\hat{D}_{k,j}^{p\,-1}
\hat{\mathrm{a}}^{\dag}_{k,j}\hat{D}_{k,j}^{p} =
\hat{\mathrm{a}}^{\dag}_{k,j} +\sum_p\frac{\hat{\bbm[\sigma]}_p \,
g_{k,j}^p}{\hbar\omega_k} \mathrm{e}^{-i  \bbm[k]\cdot
\bbm[r]_0}\,.
\end{eqnarray}
To treat properly the initial condition of particle and bath when
correlations are present at all frequencies, we apply the
canonical transformation given in Eq.\,(\ref{operatore
trasformazione}) to all the modes. This amounts to express
$\hat{H}$ in terms of the transformed operators
$\mathrm{\tilde{a}}_{k,j}$ and $ \mathrm{\tilde{a}}^{\dag}_{k,j}$.
Thus, inverting Eq.\,(\ref{trasformazione}) and substituting into
the Hamiltonian of Eq.\,(\ref{hamiltoniana}), we obtain
\begin{eqnarray}\label{H trasformata}
  \hat{H}= \sum_p \frac{p^2}{2m}\;\hat{\bbm[\sigma]}_p + \sum_{k,j}\hbar \omega_k
  \mathrm{\widetilde{a}}^{\dag}_{k,j}\mathrm{\widetilde{a}}_{k,j}
   +\sum_p \frac{p^2}{2m}\frac{\delta
  m}{m}\;\hat{\bbm[\sigma]}_p -\sum_{p,k,j}\frac{
  g_{k,j}^{p\,2}}{\hbar\omega_k}\;\hat{\bbm[\sigma]}_{p}\, .
\end{eqnarray}
In the last term only $\hat{\bbm[\sigma]}_{p}$ is an operator thus
it is possible to write
\begin{equation}\label{termine di rinormalizzazione}
 \sum_{k,j}\frac{ g_{k,j}^{p\,2}}{\hbar\omega_k} =
 \frac{p^2}{2m}\frac{\delta m}{m}\, ,
\end{equation}
by an appropriate definition of $\delta m$. In particular for the
charge-electromagnetic field case, using Eq.\,(\ref{coefficienti
di accoppiamento}) for $g_{k,j}^{p}$, taking the continuum limit,
the mass variation $\delta m$ has the form
\begin{equation}\label{mass variation}
\delta m = \frac{4\alpha \hbar\Omega}{3 \pi c^2}
\frac{m^2}{m_{0}^2} \, ,
\end{equation}
where $\Omega$ is an upper frequency cut off. We see that, within
second order perturbation theory in the charge, it coincides with
the usual electromagnetic mass variation due to the interaction
with the electromagnetic field \cite{Sakurai (1977)}.
\\
In this way, inserting Eq.\,(\ref{termine di rinormalizzazione})
in Eq.\,(\ref{H trasformata}), the Hamiltonian reduces to
\begin{equation}\label{H trasformata2}
  \hat{H}=\sum_p \frac{p^2}{2m}\;\hat{\bbm[\sigma]}_p+\sum_{k,j}\hbar \omega_k
  \mathrm{\widetilde{a}}^{\dag}_{k,j}\mathrm{\widetilde{a}}_{k,j}\,.
\end{equation}
The eigenstates of $\hat{H}$ belong to the complete Hilbert space
$\mathcal{H}_S\otimes\mathcal{H}_F$ and have the form
$|p,\{\widetilde{n}_{k,j}\}\rangle =
\prod_{k,j}\hat{D}_{k,j}^p|p\rangle |n_{k,j}\rangle$. In the case
where each momentum is in equilibrium with the dressed vacuum
state, we consider the states
\begin{eqnarray}\label{autostati}
  |\widetilde{p}\rangle =|p,\{\widetilde{0}_{k,j}\}\rangle=
   \prod_{k,j} \hat{D}_{k,j}^p|p\rangle | 0_{k,j} \rangle
  = |p\rangle \prod_{k,j}\hat{D}_{k,j}^{(p)}| 0_{k,j} \rangle
  =|p\rangle | \{\alpha_{k,j} (p)\}\rangle \, .
\end{eqnarray}
In Eq.\,(\ref{autostati}) each operator $\hat{D}_{k,j}^{(p)}$ acts
only on the field-mode Hilbert space $\mathcal{H}_F^{k,j}$.
Because of the form of $\hat{D}_{k,j}^{(p)}$, the state
$|\{\alpha_{k,j} (p)\}\rangle $ indicates the product of the
coherent states of all the modes  of the field, each of amplitude
$\alpha_{k,j} (p)= g_{k,j}^p \exp(-i  \bbm[k]\cdot \bbm[r]_0)
/\hbar\omega_k$, associated to the component $|p\rangle$ of the
wave packet in the space $\mathcal{H}_S$. The state
$|\widetilde{p}\rangle$ represents in the electromagnetic case the
state of the coupled system formed by the particle of momentum
$\bbm[p]$ plus the transverse photons associated to it.
\\
The action of Hamiltonian (\ref{H trasformata2}) on the states
$|\widetilde{p}\rangle$ of Eq.\,(\ref{autostati}) reduces to
\begin{equation}
  \hat{H}|\widetilde{p}\rangle =\frac{p^2}{2m}
  |\widetilde{p}\rangle \, .
\end{equation}
Thus $|\widetilde{p}\rangle$ are eigenstates of $\hat{H}$ with
eigenvalues  $p^2/2m$.
\\
In the Schr\"{o}dinger picture the time evolution operator
expressed in terms of the transformed operators is
\begin{eqnarray}\label{operatore evoluzione}
  &\hat{U}(t)=\exp \left[-\frac{i}{\hbar} \left(\sum_p
\frac{p^2}{2m}\;\hat{\bbm[\sigma]}_p+\sum_{k,j}\hbar \omega_k
  \mathrm{\widetilde{a}}^{\dag}_{k,j}\mathrm{\widetilde{a}}_{k,j}
   \right)\,t\right]\,.
\end{eqnarray}
 We consider an initial localized wave packet whose
components are the correlated states of the particle in
equilibrium with the bath in its vacuum state
\begin{equation}\label{initial state}
  |\Psi\rangle =\sum_{p} C_p|\widetilde{p}\rangle \,.
\end{equation}
The initial density matrix of the total system is then given by
\begin{eqnarray}\label{matrice densita iniziale}
  \widetilde{\rho} (0) =\sum_{p, p'} C_p C^{*}_{p'}|\widetilde{p}\rangle
  \langle \widetilde{p}'|   =\sum_{p, p'} C_p C^{*}_{p'}|p\rangle\langle p'| \otimes
  |\{\widetilde{0}_{k,j} \}\rangle   \langle \{\widetilde{0}_{k',j'}
  \}|\,.
\end{eqnarray}
From Eq.\,(\ref{matrice densita iniziale}) the reduced density
matrix of the particle at time $t$ is given by
\begin{eqnarray}\label{matrice densita ridotta tempo 1}
  \widetilde{\rho}_S(t)= \mathrm{tr}_F \{
  \hat{U}(t) \sum_{p, p'} C_p C^{*}_{p'}|p\rangle\langle p'|  \otimes
  |\{\widetilde{0}_{k,j} (p)\}\rangle  \langle \{\widetilde{0}_{k',j'} (p')\}|
  \hat{U}^{-1}(t)\} \,.
\end{eqnarray}
Inserting the time evolution operator (\ref{operatore evoluzione})
and the expression for $|\widetilde{p}\rangle$ (\ref{autostati})
in Eq.\,(\ref{matrice densita ridotta tempo 1}), the matrix
elements of the reduced density matrix can be cast in the form
\begin{eqnarray}\label{elemento matrice ridotta tempo 2}
  \widetilde{\rho}_S^{p, p'}(t)= & C_p C^{*}_{p'}\exp \left[{-\frac{it}{\hbar}\frac{(p^2-p'^2)}{2m}}\right]
  \nonumber \mathrm{tr}_F \left\{|\{\alpha_{k,j} (p)\}\rangle   \langle \{\alpha_{k',j'}
  (p')\}|\right\}\,.
\end{eqnarray}
The time dependence of $\widetilde{\rho}_S^{p, p'}(t)$ is given
simply by $\exp \left[-it(p^2-p'^2)/2m \hbar\right]$, and
represents the free evolution of the initial reduced matrix
elements
\begin{equation}\label{elemento matrice iniziale}
  \widetilde{\rho}_S^{p,p'}(0)=C_p C^{*}_{p'}
  \mathrm{tr}_F \left\{|\{\alpha_{k,j} (p)\}\rangle   \langle \{\alpha_{k',j'}
  (p')\}|\right\}\,.
\end{equation}
It follows that for an initial state consisting of a coherent wave
packet of particle-bath correlated states $|\widetilde{p}\rangle$,
in the form of Eq.\,(\ref{matrice densita iniziale}), decoherence
doesn't depend on time. The decoherence present in the reduced
density matrix elements is contained in the factor $\mathrm{tr}_F
\left\{|\{\alpha_{k,j} (p)\}\rangle \langle \{\alpha_{k',j'}
(p')\}|\right\}$ appearing in Eq.\,(\ref{elemento matrice
iniziale}). For charge-field interaction, this factor can be
interpreted as a consequence of the cloud of transverse virtual
photons being associated to each momentum component  of the
particle.
\\
This interpretation of the presence of decoherence in our system
is analogous to the one given for the case of two level systems
linearly coupled to a bath of harmonic oscillators
\cite{Palma-Suominen-Ekert (1996)}, where decoherence is
attributed to the buildup of correlations between the environment
and the states of the two level systems.

By performing explicitly the trace on the field in the factor
appearing in Eq.\,(\ref{elemento matrice iniziale}) we obtain
\begin{eqnarray} \label{traccia1}
\fl  \mathrm{tr}_F \left\{|\{\alpha_{k,j} (p)\}\rangle \langle
\{\alpha_{k',j'} (p')\}|\right\}& =
 \prod_{k'',j''}
\sum_n \langle n_{k'',j''}| \prod_{k,k',j,j'} \hat{D}_{k,j}^{(p)}
| 0_{k,j} \rangle \langle 0_{k',j'}| \hat{D}_{k',j'}^{(p)-1}
|n_{k'',j''}\rangle \nonumber \\ \fl &=  \prod_{k,k',j,j'} \langle
0_{k',j'}|  \exp \left[ \frac{g_{k',j'}^{p'}}{\hbar\omega_{k'}}
\left( \hat{\mathrm{a}}_{k',j'}\mathrm{e}^{i
\bbm[k]'\cdot\bbm[r]_0}
-\hat{\mathrm{a}}^{\dag}_{k',j'}\mathrm{e}^{-i
\bbm[k]'\cdot\bbm[r]_0}\right)\right] \nonumber \\ \fl  & \; \;
\;\;\; \times \exp\left[ \frac{ g_{k,j}^p}{\hbar\omega_k} \left(
\hat{\mathrm{a}}^{\dag}_{k,j}\mathrm{e}^{-i
\bbm[k]\cdot\bbm[r]_0}-\hat{\mathrm{a}}_{k,j}\mathrm{e}^{i
\bbm[k]\cdot\bbm[r]_0} \right)\right] | 0_{k,j} \rangle \nonumber
\\ \fl & =  \prod_{k,k',j,j'} \langle
\frac{g_{k',j'}^{p'}}{\hbar\omega_{k'}} |\frac{
g_{k,j}^p}{\hbar\omega_k} \rangle \,.
\end{eqnarray}
Hence, the trace reduces to a product of coherent states. We
calculate Eq.\,(\ref{traccia1}) explicitly in the case of a
charged particle interacting with the electromagnetic field. We
use the expression  for the $g_{k,j}^p$ of Eq.\,(\ref{coefficienti
di accoppiamento}), exploit the form of the scalar product between
coherent states and perform the continuum limit of the field
modes, $\sum_k \rightarrow \frac{V}{(2\pi c)^3}\int_0^\infty
\mathrm{d}^3 \omega$, to get
 \begin{eqnarray}\label{traccia2}
  \fl &\mathrm{tr}_F \left\{|\{\alpha_{k,j} (p)\}\rangle
\langle \{\alpha_{k',j'} (p')\}|\right\} = \nonumber \\ \fl &  =
\prod_{k,j}
  \exp \left[-\frac{\left(g_{k,j}^{p}-g_{k,j}^{p'}\right)^2}{2\hbar^2\omega_{k}^2}
   \right]
    =
    \prod_{k,j}
  \exp \left\{-\frac{e^2}{m_0^2}\frac{2\pi\hbar}{V
\omega_k } \frac{\left[(\bbm[p]-\bbm[p]\,')\cdot
\bbm[\varepsilon]_{k,j}\right]^2}{2\hbar^2\omega_{k}^2}
   \right\}
     \\ \fl &  =
  \exp \left\{-\sum_j\frac{e^2}{8\pi^2 c^3m_0^2\hbar}
  \int_0^\infty \mathrm{d}^3\omega
  \frac{\left[(\bbm[p]-\bbm[p]\,')\cdot
\bbm[\varepsilon]_{k,j}\right]^2}{\omega^3}
   \right\}  =
     \exp \left\{-\frac{\alpha |\bbm[p]-\bbm[p]\,'|^2}{3\pi m_0^2 c^2}
  \int_0^\infty \mathrm{d}\omega
  \frac{\mathrm{e}^{-\omega/\Omega}}{\omega}
       \right\}\nonumber \, ,
  \end{eqnarray}
where $\alpha=e^2/\hbar c$. The last equation has been obtained
summing over the polarizations and introducing as usual a high
frequency cut off factor in the integral over frequencies
\cite{Breuer-Petruccione (2001)}. However the frequency integral
appearing in the last equation maintains the infrared divergence.
This makes the reduced density matrix of the particle diagonal and
the decoherence between different momentum states of the particle
is therefore complete. Thus, as a consequence of the infrared
divergence, there is a super selection rule in the momenta with
the emergence of stable super selection sectors \cite{Kupsch
(2002)}.
\\
However, by taking into account the finite time  measurement of
the particle momenta, a lower cut off frequency $\varpi$ may be
introduced, the frequency $\varpi$ representing the resolution in
the detection process. Introducing this lower frequency cut off
$\varpi$ in Eq.\,(\ref{traccia2}) we obtain \cite{Abramowitz
(1972)}
\begin{eqnarray} \label{traccia3}
  \fl \mathrm{tr}_F \left\{|\{\alpha_{k,j} (p)\}\rangle
  \langle \{\alpha_{k',j'} (p')\}|\right\} & =
  \exp \left\{-\frac{\alpha |\bbm[p]-\bbm[p]\,'|^2}{3\pi m_0^2 c^2}
  \int_\varpi^\infty \mathrm{d}\omega
  \frac{\mathrm{e}^{-\omega/\Omega}}{\omega}
  \right\}   \\ &=
  \exp \left\{\frac{\alpha |\bbm[p]-\bbm[p]\,'|^2}{3\pi m_0^2 c^2}
  \left[\gamma +\ln \frac{\varpi}{\Omega} + \sum_{n=1}^{\infty}\frac{(-1)^n}{n\,
  n!}\,\left(\frac{\varpi}{\Omega}\right)^n \right]
  \right\} \,,\nonumber
\end{eqnarray}
where $\gamma=0.577216$ is the Euler's constant.
\\
For $\varpi/\Omega \ll 1$  Eq.\,(\ref{traccia3}) can be
approximated as
\begin{equation} \label{traccia4}
  \mathrm{tr}_F \left\{|\{\alpha_{k,j} (p)\}\rangle
  \langle \{\alpha_{k',j'} (p')\}|\right\}=
  \exp \left\{\frac{\alpha |\bbm[p]-\bbm[p]\,'|^2}{3\pi m_0^2 c^2}
  \ln \frac{\varpi}{\Omega}
  \right\}\,,
\end{equation}
and the initial reduced density matrix elements of
Eq.\,(\ref{elemento matrice iniziale}) take the form
\begin{equation}\label{elemento matrice iniziale2}
  \widetilde{\rho}_S^{p, p'}(0)= C_p C^{*}_{p'}
   \exp \left\{\frac{\alpha |\bbm[p]-\bbm[p]\,'|^2}{3\pi m_0^2 c^2}
  \ln \frac{\varpi}{\Omega}
  \right\}  \,.
\end{equation}
Now decoherence between momenta in the matrix elements of
Eq.\,(\ref{elemento matrice iniziale2}) is not complete anymore
and depends both on the distance between the values of the
particle momenta and on $\varpi$.

\section{Partially correlated states}

The need of introducing a lower cut off frequency can also derive
from the initial packet preparation rather than from the final
measurement. If the initial packet is localized by a measurement
with an uncertainty  $\Delta x$ and then with a momentum
uncertainty $\Delta p \sim \hbar / \Delta x $, this is equivalent
to a minimum measurement time $\Delta t \sim \hbar/\Delta p \Delta
v$ \cite{Petruccione-Breuer libro (2002),Wheeler-Zurek (1983)}, to
which it is possible to associate a lower frequency limit  $\varpi
\sim 1/\Delta t $ to modes that have reached equilibrium with the
particle.
\\
This indicates that we can start from an initial condition with
the particle in equilibrium only with the modes at a frequency
$\omega
> \varpi$ while it has no correlation with those at $\omega<\varpi$. Thus we
take as initial condition a state whose momentum components have
the form
\begin{equation}\label{partiallycorrelatedinitialstate}
  |\tilde{p}_\varpi\rangle =|p\rangle \prod_{j} \prod_{k=0}^{\varpi /c}
  |0_{k,j}\rangle
   \prod_{\bar{k}=\varpi /
   c}^{\infty}|\tilde{0}_{\bar{k},j}\rangle \;.
\end{equation}
From Eq.\,(\ref{partiallycorrelatedinitialstate}) we build the
initial wave packet $\sum_p C_p |\tilde{p}_\varpi\rangle$ that
corresponds to the initial density matrix
\begin{eqnarray}\label{initial density matrix}
  \tilde{\rho}_{\varpi} (0)=\sum_{p, p'} C_p C^{*}_{p'}|p\rangle\langle p'|
  \prod_{j}
   \prod_{k=0}^{\varpi /c}
  |0_{k,j}\rangle \langle 0_{k,j}|
  \prod_{\bar{k}=\varpi / c}^{\infty}|{\tilde{0}}_{\bar{k},j}\rangle
  \langle \tilde{0}_{\bar{k},j}| \; .
\end{eqnarray}
In Eq.\,(\ref{initial density matrix}) the modes at frequency
higher than $\varpi$ are distinguished from those at lower
frequency and we have eliminated the cross terms between the two
mode regions. In fact these terms won't contribute to the reduced
density matrix elements because they will be eliminated in the
field trace and then we don't consider them.

To discuss the time evolution from the initial state represented
by the  density matrix $\tilde{\rho}_{\varpi} (0) $ it is useful
to separate in the Hamiltonian of Eq.\,(\ref{hamiltoniana}) the
parts at frequency larger and smaller than $\varpi$, i.e.
\begin{eqnarray}\label{hamiltoniana3}
  \fl \hat{H}&=\sum_p
\frac{p^2}{2m}\;\hat{\bbm[\sigma]}_p+\hat{H}_{\omega<\varpi} +
\hat{H}_{\omega
> \varpi}
+\sum_p \frac{p^2}{2m}\frac{\delta m}{m}\;\hat{\bbm[\sigma]}_p \nonumber \\
\fl &=  \sum_p \frac{p^2}{2m}\;\hat{\bbm[\sigma]}_p
 +\left[ \sum_{j,k=0}^{\varpi/c}\hbar \omega_k
  \hat{\mathrm{a}}^{\dag}_{k,j}\hat{\mathrm{a}}_{k,j}
  + \sum_{p,j,k=0}^{\varpi/c}\hat{\bbm[\sigma]}_p \, g_{k,j}^p\left(
  \hat{\mathrm{a}}^{\dag}_{k,j}\mathrm{e}^{-i \bbm[k]\cdot\bbm[r]_0}+\hat{\mathrm{a}}_{k,j}\mathrm{e}^{i
  \bbm[k]\cdot\bbm[r]_0} \right) \right]  \\ \fl &
  + \left[ \sum_{j,k=\varpi/c}^{\infty}\hbar \omega_k
  \hat{\mathrm{a}}^{\dag}_{k,j}\hat{\mathrm{a}}_{k,j}
  +\sum_{p,j,k=\varpi/c}^{\infty}\hat{\bbm[\sigma]}_p \, g_{k,j}^p\left(
  \hat{\mathrm{a}}^{\dag}_{k,j}\mathrm{e}^{-i \bbm[k]\cdot\bbm[r]_0}+\hat{\mathrm{a}}_{k,j}\mathrm{e}^{i
  \bbm[k]\cdot\bbm[r]_0}\right)   \right]
  +\sum_p  \frac{p^2}{2m}\frac{\delta m}{m}\;\hat{\bbm[\sigma]}_p \; . \nonumber
\end{eqnarray}
To treat  properly  the presence in the initial condition of
correlations with the high frequency modes, in $\hat{H}$, we apply
the canonical transformation (\ref{operatore trasformazione}) only
to the modes at frequency higher than $\varpi$. This amounts to
express  only $\hat{H}_{\omega
> \varpi}$ in terms of the transformed
operators $\mathrm{\tilde{a}}_{k,j}$ and $
\mathrm{\tilde{a}}^{\dag}_{k,j}$. Following the same steps leading
from Eq.\,(\ref{hamiltoniana}) to Eq.\,(\ref{H trasformata}) we
get
\begin{eqnarray}\label{H trasformata3}
  \fl \hat{H}=\sum_p
\frac{p^2}{2m}\;\hat{\bbm[\sigma]}_p+\hat{H}_{\omega<\varpi} +
  \sum_{j,k=\varpi/c}^\infty \hbar \omega_k
  \mathrm{\tilde{a}}^{\dag}_{k,j}\mathrm{\tilde{a}}_{k,j}
  +\sum_p
  \frac{p^2}{2m}\frac{\delta m}{m}\;\hat{\bbm[\sigma]}_p -\sum_{p,j,k=\varpi/c}^\infty \frac{
  g_{k,j}^{p\,2}}{\hbar\omega_k}\hat{\bbm[\sigma]}_{p}  \, .
\end{eqnarray}
In the last term of Eq.\,(\ref{H trasformata3}) only
$\hat{\bbm[\sigma]}_{p}$ is an operator and, in analogy to
Eq.\,(\ref{termine di rinormalizzazione}), it is possible to write
by an appropriate definition of $\delta m_{>\varpi}$
\begin{eqnarray}\label{termine di rinormalizzazione2}
 \sum_{j,k=\varpi/c}^\infty \frac{ g_{k,j}^{p\,2}}{\hbar\omega_k} =
 \frac{p^2}{2m}\frac{\delta m_{>\varpi}}{m}\, .
\end{eqnarray}
$\delta m_{>\varpi}$ corresponds to the mass variation of the
particle due to dressing with the high frequency modes. Thus, for
the sum of the last two terms present in Eq.\,(\ref{H
trasformata3}) we obtain
\begin{eqnarray}\label{nuova H con S}
  \sum_p
  \frac{p^2}{2m}\frac{\delta m}{m}\;\hat{\bbm[\sigma]}_p-\sum_{p,j,k=\varpi/c}^\infty \frac{
  g_{k,j}^{p\,2}}{\hbar\omega_k}\;\hat{\bbm[\sigma]}_p= \sum_p \frac{p^2}{2m}\frac{\delta m_{<\varpi}}{m}\;\hat{\bbm[\sigma]}_p \; ,
\end{eqnarray}
where $\delta m_{<\varpi}=\delta m-\delta m_{>\varpi}$ indicates
the mass variation of the particle due to dressing with the low
frequency modes. Using Eq.\,(\ref{nuova H con S}) in Eq.\,(\ref{H
trasformata3}), the Hamiltonian $\hat{H}$ becomes
\begin{eqnarray}\label{H trasformata4}
  \fl \hat{H}  =\sum_p
  \frac{p^2}{2m}\;\hat{\bbm[\sigma]}_p + \sum_p \frac{p^2}{2m}\frac{\delta m_{<\varpi}}{m}\;\hat{\bbm[\sigma]}_p+
  \hat{H}_{\omega<\varpi} +
  \tilde{H}_{\omega>\varpi}  =
  \sum_p \frac{p^2}{2m_{\varpi}}\, \hat{\bbm[\sigma]}_p +\hat{H}_{\omega<\varpi} +
  \tilde{H}_{\omega>\varpi}   \;,
\end{eqnarray}
where $\tilde{H}_{\omega>\varpi}=\sum_{j,k=\varpi/c}^\infty \hbar
\omega_k \mathrm{\tilde{a}}^{\dag}_{k,j}\mathrm{\tilde{a}}_{k,j}$
and the mass  $m_{\varpi}$ is defined by
\begin{equation}\label{massa di omega}
  \frac{1}{m_{\varpi}}= \frac{1}{m}\left(1+\frac{\delta
m_{<\varpi}}{m}\right)\,.
\end{equation}
No renormalization term is present in the Hamiltonian of
Eq.\,(\ref{H trasformata4}). This is due to the fact that only the
interaction with modes at frequency $\omega<\varpi$ appears in $H$
and that the mass $m_{\varpi}$ represents the  bare mass with
respect to the low frequency modes. The Hamiltonian being now
expressed in terms of the bare mass, the low frequency
renormalization term disappears.

In the Hamiltonian $\hat{H}$ (\ref{H trasformata4}) the terms $
\sum_p ( p^2  /2m_{\varpi}) \, \hat{\bbm[\sigma]}_p $,
$\hat{H}_{\omega<\varpi}$ and $\tilde{H}_{\omega>\varpi}$ commute
among them. This allows to analyze in a transparent way the
influence of each of these terms on the time evolution of the
reduced density matrix  elements of $\tilde{\rho}_{\varpi} (0)$
and to discuss separately the effects of  high frequency
correlated and low frequency uncorrelated modes. In fact we can
write the reduced density matrix elements of the particle at time
$t$ as
\begin{equation}\label{elementi matrice ridotta}
  \tilde{\rho}_{S\,\varpi}^{p, p'}(t)=\langle
  p|\mathrm{tr}_F \{\hat{U}_{\varpi}(t) \tilde{\rho}_{\varpi} (0)
  \hat{U}_{\varpi}^{-1}(t)\}|p'\rangle \; ,
\end{equation}
where $\hat{U}_{\varpi}(t)$ is the time evolution operator, that
may be written as the product of three operators
$\hat{U}_{\varpi}(t)=\hat{U}_{S,\varpi}(t)\hat{U}_{\omega<\varpi}(t)\tilde{U}_{\omega>\varpi}(t)$.
The first,  $\hat{U}_{S,\varpi}=\exp \left [-i \sum_p
(p^2/2m_{\varpi} \hbar)\hat{\bbm[\sigma]}_p \,t\right]$, depends
on the kinetic energy and can be shown to affect only the free
evolution of reduced density matrix elements. Introducing the
time-ordering operator $T_{\leftarrow}$, the second term is $
\hat{U}_{\omega<\varpi}(t)= \mathrm{T}_{\leftarrow}\exp \left
[-i\int_0^t \mathrm{d}s \hat{H}_{\omega<\varpi} (s)/\hbar
\right]$, and depends on the modes at frequency lower than
$\varpi$, while the third term is $\tilde{U}_{\omega>\varpi}(t)=
\mathrm{T}_{\leftarrow}\exp \left [-i\int_0^t \mathrm{d}s
\tilde{H}_{\omega>\varpi}(s) /\hbar\right]$ and depends on those
at frequency higher than $\varpi$. Thus, Eq.\,(\ref{elementi
matrice ridotta}) can be written as
\begin{eqnarray}\label{matrice densita semivestita formale}
  \fl \tilde{\rho}_{S\,\varpi}^{p, p'}(t)  =
   \exp
  \left[-\frac{it}{\hbar}\frac{(p^2-p'^2)}{2m_{\varpi}}\right]
   \langle
  p|\mathrm{tr}_F \{\hat{U}_{\omega<\varpi}(t)
  \left[\tilde{U}_{\omega>\varpi}(t) \tilde{\rho}_{\varpi} (0)
  \tilde{U}_{\omega>\varpi}^{-1}(t)\right]
  \hat{U}_{\omega<\varpi}^{-1}(t)\}|p'\rangle \; .
\end{eqnarray}
$\hat{U}_{\omega<\varpi}$ acts only on the low frequency part of
the density operator $\tilde{\rho}_{\varpi} (0)$ of
Eq.\,(\ref{initial density matrix}) representing the modes
uncorrelated to the particle momenta. This leads to a time
evolution of this part of $\tilde{\rho}_{\varpi} (0)$ formally
identical to the one we would get from a fully uncorrelated
initial condition.

For a charge interacting with the electromagnetic field  we may
directly use, for the evolution of the low frequency part of
$\tilde{\rho}_{\varpi} (t)$, the results already obtained in
\cite{bellomo (2004)}, that is:
\begin{equation}\label{funzione di decoerenza}
  \rho_S^{p,p'}(t)=\rho_S^{p,p'}(0)\exp\left[-\frac{it}{\hbar}\frac{(p^2-p'^2)}{2m_0}\right]
  \exp{\Gamma^{p,p'}(t)} \; ,
\end{equation}
where the complex decoherence function
$\Gamma^{p,p'}(t)=\Gamma^{p,p'}_r(t)+i\Gamma^{p,p'}_i(t)$  has
been introduced with $\Gamma^{p,p'}_r(t)$ and $\Gamma^{p,p'}_i(t)$
given respectively by the integral expressions
\begin{eqnarray}\label{contributo reale}
  \Gamma^{p,p'}_{r}(t)=-\frac{2\alpha }{3\pi}\frac{|\bbm[p]-\bbm[p]\,'|^2}{m_0^2c^2}
  \int_0^\infty \mathrm{d}
  \omega \exp\left(-\frac{\omega}{\Omega}\right) \frac{(1-\cos \omega t)
  }{\omega}   \; ,
\end{eqnarray}
which corresponds to the expression of  $\Gamma^{p,p'}_{r}(t)$
given in \cite{bellomo (2004)} in the limit $T=0$, in which only
the vacuum contribution remains so that
$\Gamma^{p,p'}_{r}(t)=\Gamma^{p,p'}_{vac}(t)$, and
\begin{eqnarray}\label{fattore di fase}
  \Gamma^{p,p'}_i(t)&
  =\frac{2 e^2 \left(p^2- p'^2 \right)}{3 \pi \hbar m_0^2 c^3 } \int_0^\infty \mathrm{d}\omega
  \exp\left(-\frac{\omega}{\Omega}\right)
  \left( t-\frac{\sin\omega t}{\omega }\right).
\end{eqnarray}
Here, because of Eq.\,(\ref{initial density matrix}), in the
integrals defining $\Gamma^{p,p'}_i(t)$ and $\Gamma^{p,p'}_{r}(t)$
we replace the cut off factor $\exp\left(-\omega /\Omega \right)$
with the step function $\Theta ( \varpi -\omega)$. In such a way
we obtain a new expression for the real and imaginary part of the
decoherence function, $\bar{\Gamma}^{p,p'}_{vac}(t)$ and
$\bar{\Gamma}^{p,p'}_i(t)$, which will depend only on the modes at
frequency lower than $\varpi$ as:
\begin{eqnarray}\label{contributo di vuoto}
  \fl \bar{\Gamma}^{p,p'}_{vac}(t)&=-\frac{2\alpha }{3\pi}\frac{|\bbm[p]-\bbm[p]\,'|^2}{m_0^2c^2}
  \int_0^\infty \mathrm{d}
  \omega \frac{(1-\cos \omega t)
  }{\omega} \Theta
( \varpi -\omega)
  \nonumber \\ \fl & =-\frac{2\alpha }{3\pi}\frac{|\bbm[p]-\bbm[p]\,'|^2}{m_0^2c^2}
  \left[\gamma -\mathrm{Ci}(\varpi t) + \ln \varpi t\right]  \; ,
\end{eqnarray}
\begin{eqnarray}\label{phase factor}
  \fl \bar{\Gamma}^{p,p'}_i(t) &=\frac{2 e^2 \left(p^2- p'^2 \right)}{3 \pi \hbar m_0^2 c^3 }
  \int_0^\infty \mathrm{d}\omega
  \left( t-\frac{\sin\omega t}{\omega }\right) \Theta
( \varpi -\omega) =\frac{2\alpha}{3\pi}\frac{p^2- p'^2}{m_0^2
c^2}[\varpi t -\mathrm{Si}(\varpi t)] \; ,
\end{eqnarray}
where $\mathrm{Ci}(\varpi t)$  and $\mathrm{Si}(\varpi t)$ are the
cosine and sine integral function \cite{Abramowitz (1972)}.
\\
Using the expression of $\mathrm{Ci}(\varpi t)$ for small and
large values of $\varpi t$ \cite{Abramowitz (1972)} we obtain for
$\bar{\Gamma}^{p,p'}_{vac}(t)$
\begin{equation}\label{contributo di vuoto2}
\label{cases} \bar{\Gamma}^{p,p'}_{vac}(t)\:\approx\:\cases{
-\frac{2\alpha}{3\pi}\frac{|\bbm[p]-\bbm[p]\,'|^2}{m_0^2c^2}\frac{\varpi^2
t^2}{4}&for $\varpi t\ll 1$\\
-\frac{2\alpha
  }{3\pi}\frac{|\bbm[p]-\bbm[p]\,'|^2}{m_0^2c^2}
  \ln \varpi t &for $\varpi t\gg  1$\\}
\end{equation}
Eq.\,(\ref{contributo di vuoto2}) shows that there are a quadratic
and a logarithmic time evolution regimes. The transition from the
first to the second one occurs at a typical time $\tilde{t}\approx
\varpi^{-1}$ with $|\bar{\Gamma}^{p,p'}_{vac}(\tilde{t})|=Q
\approx
\frac{2\alpha}{3\pi}\frac{|\bbm[p]-\bbm[p]\,'|^2}{m_0^2c^2}$. The
effective magnitude of $\bar{\Gamma}^{p,p'}_{vac}(t)$ at a given
time depends both on $\varpi$, which is determined by the
preparation or observation procedure, and on $Q$ which depends on
the charge and the mass of the particle and on the distance of the
reduced density matrix element from the diagonal in the momentum
space. The vacuum influence on the time evolution of decoherence
for $Q\ll 1$ remains small because of the logarithmic dependence
of $\bar{\Gamma}^{p,p'}_{vac}(t)$ from time. Its effect increases
with $Q$ and for sufficiently large $Q$ decoherence may become
effective at  times $t \geq \varpi^{-1}$. This behavior is shown
in figure 1.
\begin{figure}[h]\begin{center}\label{grafico}
\includegraphics[width=0.7\textwidth]{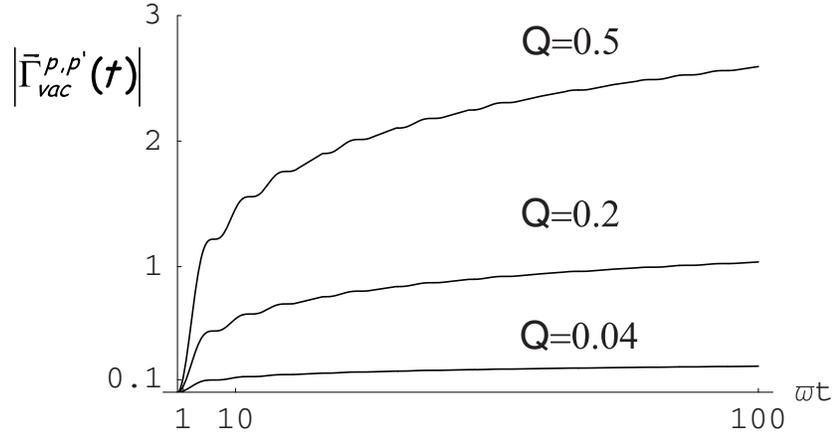}
\caption{Time dependence of $|\bar{\Gamma}^{p,p'}_{vac}(t)|$ for
different values of the parameter $Q$.}
\end{center}\end{figure}
\\
In conclusion we see that the effect of the low frequency modes,
described by $U_{\omega<\varpi}(t)$, reduces to a multiplication
factor in the reduced density matrix element
$\tilde{\rho}_{S\,\varpi}^{p, p'}(t)$ of the form $ \exp
\left[{\bar{\Gamma}^{p,p'}_{vac}(t)+i \bar{\Gamma}^{p,p'}_i(t) }
\right]$ where $\bar{\Gamma}^{p,p'}_{vac}(t)$ is given by
Eq.\,(\ref{contributo di vuoto})
 and $\bar{\Gamma}^{p,p'}_i(t)$ by Eq.\,(\ref{phase factor}).

With regards to the third term in the Hamiltonian (\ref{H
trasformata4}), $\tilde{H}_{\omega>\varpi}$, representing the high
frequency modes contribution, its action on the time evolution of
$\tilde{\rho}_{S\,\varpi}^{p, p'}(t)$ can be obtained by following
the equations that from Eq.\,(\ref{operatore evoluzione}) lead to
Eq.\,(\ref{elemento matrice iniziale2}), where the lower frequency
represents in that case the effect of the finite resolution of the
detection process. We are therefore led to a term formally
identical with the one of Eq.\,(\ref{elemento matrice iniziale2})
that doesn't influence the low frequency part of
$\tilde{\rho}_{\varpi} (0)$.

Joining together all the previous  considerations we finally
obtain for the complete expression of the reduced density matrix
elements
\begin{eqnarray}\label{matrice densita in considerazione semivestita}
  \fl\tilde{\rho}_{S\,\varpi}^{p, p'}(t)= & C_p C^{*}_{p'}
  \exp
  \left[-\frac{it}{\hbar}\frac{(p^2-p'^2)}{2m_{\varpi}}\right]
   \exp \left[ i  \frac{2\alpha \left(\varpi t -\mathrm{Si}(\varpi t)\right)}
   {3 \pi   m_0^2 c^2}
  (p^2-p'^2)\right]
   \nonumber \\\fl &\left.\times \exp \left\{
  - \frac{2 \, \alpha \, |\bbm[p]-\bbm[p]\,'|^2}{3 \pi m_0^2 c^2}
  \left[\gamma -\mathrm{Ci}(\varpi t) + \ln \varpi t\right] \right\} \right.
   \exp \left[\frac{\alpha |\bbm[p]-\bbm[p]\,'|^2}{3\pi m_0^2 c^2}
  \ln \frac{\varpi}{\Omega}
  \right] \; .
\end{eqnarray}
We observe that with regard to the contribution in
Eq.\,(\ref{matrice densita in considerazione semivestita}) of the
low frequency modes, we have used Eqs.\,(\ref{contributo di
vuoto}) and (\ref{phase factor}) whose form is valid at all the
times.  For the contribution of the high frequency modes we have
used Eq.\,(\ref{elemento matrice iniziale2}) which is valid in the
limit $\varpi/\Omega \ll 1$. However it is possible from
Eq.\,(\ref{matrice densita in considerazione semivestita}) to
obtain a simplified form of $\tilde{\rho}_{S\,\varpi}^{p, p'}(t)$
for small and large $\varpi t$, using for
$\bar{\Gamma}^{p,p'}_{vac}(t)$ the corresponding expressions of
Eq.\,(\ref{contributo di vuoto2}).

The form given by Eq.\,(\ref{matrice densita in considerazione
semivestita}) describes the time evolution of the reduced density
matrix elements when, in the initial state, correlations with the
field modes at frequency larger than $\varpi$ are present. The
form for $\tilde{\rho}_{S\,\varpi}^{p, p'}(t)$ of
Eq.\,(\ref{matrice densita in considerazione semivestita}) is
intermediate between the two extreme forms, completely correlated
(\ref{elemento matrice ridotta tempo 2}, \ref{traccia2}) and
uncorrelated (\ref{funzione di decoerenza}), and it coincides with
them respectively in the limits $\varpi\rightarrow 0$ and
$\varpi\rightarrow \Omega$.

One must observe that in the fully  correlated case ($\varpi =0$)
the particle is completely entangled with the field and the
initial reduced density matrix (\ref{matrice densita iniziale})
contains already all the decoherence, which therefore remains
constant with time. For $\varpi \neq 0$ the initial entanglement
is partial and it is the coupling dynamics that induces the
progressive entanglement with the modes at frequency less than
$\varpi$ thus leading to an increase of the decoherence present at
$t=0$.

\section{Summary and Conclusions}

We have studied the decoherence among the momentum components of a
free particle wave packet linearly interacting with a zero
temperature bath by working out the time dependence of the off
diagonal elements of the reduced density matrix of the particle.
We have also examined the influence on the decoherence development
by the initial conditions and in particular by the presence of
correlations between the particle and the environment.

The influence of initial conditions on decoherence has been
previously studied by examining the attenuation of coherence
between a pair of coherent wave packets in the case of a particle
interacting with a bath either at finite \cite{Lutz (2003)} or
zero temperature \cite{Ford-O'Connell (2003)}.

We have considered initial conditions ranging from the case of
absence of initial correlations to the case where they are fully
developed. The particle-bath system is described in the former
case by a factorized density matrix and in the latter, which
corresponds to a stationary condition, by a linear combination of
the eigenstates of the total Hamiltonian. The reason to consider
this range of initial conditions is linked to the fact that when
the interaction between the system and the bath is always present,
as in the case of a charged particle interacting with the
radiation field, it isn't appropriate to start from a condition
where no correlations are present. On the other hand because of
either a finite measurement time or preparation time it is not
appropriate to start from a condition where there is complete
equilibrium among the particle momenta and all the field modes. By
taking for example a finite preparation time $\tau$, one is led to
consider a situation where equilibrium correlations have been
established only with the modes at frequency higher than $\varpi
\approx 1/\tau$, while modes of lower frequency are uncorrelated.

This intermediate initial situation may be described by a density
matrix formed by two parts: the first, time independent, describes
the initial correlations already established with modes at
frequency higher than $\varpi$; the second time dependent, will
describe the establishing of correlations with the modes at
frequency lower than $\varpi$. We have shown that, for the case of
linear interaction, the increase of decoherence is at first rapid
and goes like $t^2$ while, after a transition time of the order of
$\varpi^{-1}$, slows and goes as $\ln t$, in a certain sense
reaching a plateau at $t \approx \varpi^{-1}$. The effect of
partially correlated initial conditions does show only in the
transition time, while the value that decoherence reaches at the
transition time is only a function of the physical parameters of
the particle like the value of the coupling constant, the particle
mass and the distance of the matrix element from the diagonal.

We have explicitly obtained our results for the case of a charged
particle interacting with the electromagnetic field. In fact, by
an appropriate choice of the width of the particle wave packet we
have shown that it is possible to adopt the dipole approximation.
In this context and by considering the case of low field
intensities, we have shown that the interaction Hamiltonian
becomes linear. In this case then the value of the decoherence at
the transition time $\varpi^{-1}$ depends from the combination of
parameters given by $Q \approx
\frac{2\alpha}{3\pi}\frac{|\bbm[p]-\bbm[p]\,'|^2}{m_0^2c^2}$.
Moreover the time independent part of the reduced density matrix
$\tilde{\rho}_{\varpi} (0)$ given by Eq.\,(\ref{initial density
matrix}) presents an infrared divergence when the preparation time
is large and thus $\varpi \rightarrow 0$ (\ref{elemento matrice
iniziale2}). In this case the decoherence in momentum space is
complete and the divergence is connected with the establishing of
a complete correlation with the low frequency modes. For finite
$\varpi$, with the buildup in time of correlations with the low
frequency modes these become populated. This process gives rise to
a cloud of photons around the particle which are then responsible
of mass renormalization. The two phenomena of decoherence due to
zero point modes and dressing of the particle are thus related.
Moreover when one performs an experiment that takes a finite time,
that shows the presence of decoherence, like in interference
experiments \cite{Myatt (2000), Hackermüller (2004)}, it is
appropriate to ask the form of the reduced density matrix that
correctly describes the experiment. From what said previously it
appears that the trace must be performed only on those field modes
where supposedly equilibrium hasn't been reached. Then the
characteristic time in which decoherence becomes complete is a
factor $\Omega / \varpi$ larger than the characteristic
decoherence time one obtains starting from completely uncorrelated
initial condition \cite{bellomo (2004)}.

Our analysis has been conducted in the context of non relativistic
QED which is an effective low energy theory with the cut off
frequency $\Omega$, in the spirit of modern quantum field theory,
parameterizing the physics due to the higher frequencies \cite{Zee
(2003)}. For this reason our final results must show a dependence
on $\Omega$, that is however as usual weak (logarithmic), as for
example in the case of non relativistic expression for the Lamb
shift. The appearance of the bare mass $m_0$ in our results, e.g.
in Eq.\,(\ref{elemento matrice iniziale}) and (\ref{matrice
densita in considerazione semivestita}), is due to its use in the
definition of the coupling coefficients $g_{k,j}^p$
(\ref{coefficienti di accoppiamento}). The $g_{k,j}^p$ in
Eq.\,(\ref{hamiltoniana}) and (\ref{hamiltoniana3}) could be
expressed in terms of the fully and partially dressed mass by
introducing a further renormalization term which is however
$O(e^3)$ and therefore typically neglected. This procedure would
eventually lead to final reduced density matrix expressions given
in terms of the dressed masses with a correction in the exponents
of $O(e^4)$ with respect to the actually results. This may be
alternatively seen by expressing, directly in the final
expressions for decoherence, the bare mass in terms of the dressed
masses obtaining a further term containing the product of the mass
variation of $O(e^2)$ for a factor $\alpha$.

As a final consideration we observe that a state describing an
electron not in complete equilibrium with its surrounding field
was suggested by Feinberg \cite{Feinberg (1980)} to represent the
electron immediately after a scattering event of duration $\tau$.
In fact, due to causality requirements the electron can reach
equilibrium only with its field within a region of size $l \leq c
\tau$. Similar states of incomplete equilibrium have been also
studied in QED when rapid changes occur in atomic or molecular
sources \cite{Compagno-Passante-Persico (1995)}. However the most
promising field where states of not complete equilibrium may be
implemented is solid state physics. The use of ultra fast
spectroscopic techniques has in fact permitted the generation of
almost bare electron-hole states and to follow the time evolution
of the dressing process due to the interaction with the phonon
field \cite{Huber-et al. (2001)}. Taking into account the previous
considerations, one can envisage a process where, starting from
states of not complete equilibrium, the evolution of decoherence
gives rise, in principle, to observable effects. Let's in fact
consider the scattering of a charged particle from two scattering
centers, the process lasting a finite time $\tau \approx
\varpi^{-1}$. From the considerations developed in this paper one
must expect that, after the scattering, the state of the particle
may be described by Eq.\,(\ref{initial density matrix}) and the
decoherence which develops after the scattering gives rise to a
decrease of the interference effects which depends on $\varpi$.

\end{document}